\documentclass[11pt]{article}

\def\leti{Lense--Thirring}
\def\zone{the error due to the even zonal harmonics of geopotential\ }

\def\rfr#1{(\ref{#1})}

\def\eqi{\begin{equation}}
\def\eqf{\end{equation}}
\def\eqia{\begin{eqnarray}}
\def\eqfa{\end{eqnarray}}
\def\ct#1{\cite{#1}}
\def\lb#1{\label{#1}}

\begin{document}

\noindent{\bf \LARGE{Towards a few-percent measurement of the
Lense-Thirring effect with the LAGEOS and LAGEOS II satellites?}}
\\
\\
\\
 Lorenzo Iorio\\Dipartimento Interateneo di Fisica dell'
Universit${\rm \grave{a}}$ di Bari
\\Via Amendola 173, 70126\\Bari, Italy\\
e-mail: lorenzo.iorio@libero.it

\begin{abstract}
Up to now attempts to measure the general relativistic
Lense-Thirring effect in the gravitational field of Earth have
been performed by analyzing a suitable $J_2-J_4-$free combination
of the nodes $\Omega$ of LAGEOS and LAGEOS II and the perigee
$\omega$ of LAGEOS II with the Satellite Laser Ranging technique.
The claimed total accuracy is of the order of $20$-$30\%$, but,
according to some scientists, it could be an optimistic estimate.
The main sources of systematic errors are the mismodelling in the
even zonal harmonic coefficients $J_l$ of the multipolar expansion
of the gravitational potential of Earth and the non-gravitational
perturbations which plague especially the perigee of LAGEOS II and
whose impact on the proposed measurement is difficult to be
reliably assessed. Here we present some evaluations of the
accuracy which could be reached with a different $J_2-$free
observable built up with the nodes of LAGEOS and LAGEOS II in view
of the new preliminary 2nd-generation Earth gravity models from
the GRACE mission. According to the GRACE-only based
EIGEN-GRACE02S solution, a 1-sigma upper bound of 4$\%$ for the
systematic error due to the even zonal harmonics can be obtained.
In the near future it could be possible to perform a reliable
measurement of the Lense-Thirring effect by means of the existing
LAGEOS satellites with an accuracy of a few percent by adopting a
time span of a few years. The choice of a not too long
observational temporal interval would be helpful in reducing the
impact of the secular variations of the uncancelled even zonal
harmonics $\dot J_4$ and $\dot J_6$ whose impact is difficult to
be reliably evaluated.
\end{abstract}


\section{Introduction}The general
relativistic gravitomagnetic force \cite{ciuwhe95} induced by the
gravitational field of a central rotating body of mass $M$ and
proper angular momentum $J$ is still awaiting for a direct,
unquestionable measurement. Up to now there exist some indirect
evidences of its existence as predicted by the General Theory of
Relativity (GTR in the following) in an astrophysical,
strong-field context \cite{rufsig03} and, in the weak-field and
slow-motion approximation valid throughout the Solar System, in
the fitting of the ranging data to the orbit of Moon with the
Lunar Laser Ranging (LLR) technique \cite{nor03}. The measurement
of the gravitomagnetic Schiff precession of the spins of four
spaceborne gyroscopes \cite{sch60} in the gravitational field of
Earth is the goal of the Stanford GP-B mission \cite{eveetal01}
which has been launched on April 2004. The obtainable accuracy
should be of the order of 1$\%$ or better.

The Lense-Thirring effect on the geodesic path of a test particle
freely falling in the gravitational field of a central rotating
body \cite{let18} consists of tiny secular precessions of the
longitude of the ascending node $\Omega$ and the argument of
pericentre $\omega$ of the orbit of the test particle
\begin{equation}
\dot\Omega_{\rm LT} =\frac{2GJ}{c^2 a^3(1-e^2)^{\frac{3}{2}}},\
\dot\omega_{\rm LT} =-\frac{6GJ\cos i}{c^2
a^3(1-e^2)^{\frac{3}{2}}},
\end{equation} where $a,\ e$ and $i$ are the semimajor axis, the eccentricity and the inclination, respectively, of
the orbit, $c$ is the speed of light and $G$ is the Newtonian
gravitational constant.

The LAGEOS III/LARES mission \cite{ciu86, ioretal02} was
specifically designed in order to measure such effect, but, up to
now, in spite of its scientific validity and relatively low cost,
it has not yet been approved by any space agency or scientific
institution. Recently, a drag-free version of this project
\cite{ioretal03}, in the context of the relativistic OPTIS mission
\cite{lammetal01}, is currently under examination by the German
Space Agency (DLR).
%----------------------------------------------------------
\section{The current LAGEOS-LAGEOS II Lense-Thirring experiment}
Up to now, attempts to observationally check the Lense-Thirring
effect in the gravitational field of Earth have been performed by
analyzing the accurately recovered orbits of the existing LAGEOS
and LAGEOS II satellites with the Satellite Laser Ranging
technique (SLR) \cite{ciuetal98, ciu02}. The adopted observable is
\begin{equation} \delta\dot\Omega^{\rm L
}+0.295\delta\dot\Omega^{\rm L\ II}-0.35\delta\dot\omega^{\rm L\
II}\sim 60.2\mu_{\rm LT},\label{ciufform}
 \end{equation} where
$\delta\dot\Omega$ and $\delta\dot\omega$ are the orbital
residuals of the rates of the node and the perigee and $\mu_{\rm
LT}$ is the solved--for least square parameter which is 0 in
Newtonian mechanics and 1 in GTR. The Lense-Thirring signature,
entirely adsorbed in the residuals of $\Omega$ and $\omega$
because the gravitomagnetic force has been purposely set equal to
zero in the force models, is a linear trend with a slope of 60.2
milliarcseconds per year (mas yr$^{-1}$ in the following). The
standard, statistical error is evaluated as 2$\%$. The claimed
total accuracy, including various sources of systematic errors, is
of the order of \cite{ciuetal98, ciu02} $20-30\%$. However, such
estimate would be too optimistic according to some scientists who
propose a different error budget \cite{rieetal98}.
%------------------------------------------------------------
\subsection{The error budget}
 The main
sources of systematic errors in this experiment are the
unavoidable aliasing effect due to the mismodelling in the
classical secular precessions induced on $\Omega$ and $\omega$ by
the even zonal coefficients $J_{l}$ of the multipolar expansion of
geopotential and the non--gravitational perturbations, which
severly affect the perigee of LAGEOS II \cite{luc01, luc02, luc03,
luc04, lucetal04}, whose impact on the proposed measurement is
difficult to be reliably assessed. It turns out that the
mismodelled classical precessions due to the first two even zonal
harmonics of geopotential $J_2$ and $J_4$ are the most insidious
source of error for the Lense--Thirring measurement with LAGEOS
and LAGEOS II. The combination \rfr{ciufform} is, by construction,
insensitive just to $J_2$ and $J_4$. According to the full
covariance matrix of the EGM96 gravity model \cite{lemetal98}, the
error due to the remaining uncancelled even zonal harmonics
amounts to almost 13$\%$ \cite{iorcelmec03} (1-sigma calculation).
However, if the correlations among the even zonal harmonic
coefficients are neglected and the variance matrix is used in a
1-sigma Root--Sum--Square fashion, \zone amounts to 46.6$\%$
\cite{iorcelmec03}.
%With this estimate the total error of the
%LAGEOS--LAGEOS II
%\leti\ experiment would be of the order of 50$\%$.
Such approach is
considered more realistic by some authors \cite{rieetal98} because
nothing assures that the correlations among the even zonal
harmonics of the covariance matrix of the EGM96 model, which has
been obtained during a multidecadal time span, would be the same
during an arbitrary past or future time span of a few years as
that used in the LAGEOS--LAGEOS II
\leti\ experiment.
A 1-sigma upper bound of almost 83$\%$ for the gravitational error
can be obtained by adding the absolute values of the individual
errors \cite{iorcelmec03}. Another point to be emphasized is that
the use of the perigee of LAGEOS II forces to adopt an
observational time span of many years in order to view certain
long--period harmonic perturbations of gravitational and
non--gravitational origin as empirically fitted quantities which
can be removed from the time series without corrupting the
extraction of the genuine relativistic secular trend. Indeed, it
turns out that the perigee of LAGEOS II is affected by the ocean
tidal perturbation $K_1,\ l=3,\ p=1$, which has a period of 5
years \cite{iorcelmec01}, and by a direct solar radiation pressure
harmonic with a period of 11.6 years \cite{luc01}. According to
\cite{luc01, luc02}, the non--gravitational part of the error
budget would amount to 28--$30\%$ over seven years. Moreover,
recent reexaminations of certain nonconservative accelerations
acting upon LAGEOS II would suggest that it could be reduced down
to a 13$\%$ level \cite{lucpriv} over the same time span. However,
it must be pointed out that such estimates are based on certain
refinements of the non--gravitational force models which were not
included \cite{lucetal04} in the GEODYN II orbit processor used at
the time of the analysis of \cite{ciuetal98}, especially as far as
certain tiny non--gravitational perturbations of thermal origin
\cite{luc02} are concerned. Moreover, it must also be recognized
that the estimates of the authors of \cite{rieetal98} are
different from such evaluations; indeed, it can be argued that
their evaluation of the impact of the nonconservative
accelerations on the measurement of the Lense--Thirring effect
with the perigee of LAGEOS II reported in \cite{ciuetal98} is of
the order of 48--99$\%$, if the optimistic 13$\%$ error, based on
the EGM96 full covariance, is adopted.
%-----------------------------------------------------------------
\section{The role of the new Earth gravity models from the CHAMP and GRACE missions}
From the previous considerations it could be argued that, in order
to have a rather precise and reliable estimate of the total
systematic error in the measurement of the Lense--Thirring effect
with the LAGEOS satellites it would be better to reduce the impact
of geopotential in the error budget and/or discard the perigee of
LAGEOS II which is very difficult to handle and is a relevant
source of uncertainty due to its great sensitivity to many
non--gravitational perturbations.

The forthcoming more accurate Earth gravity models from CHAMP
\cite{pav00} and, especially, GRACE \cite{riesetal03} will yield
an opportunity to realize both these goals, at least to a certain
extent.
%---------------------------------------------------------------------
\subsection{The EIGEN-GRACE02S model}
In order to evaluate quantitatively the opportunities offered by
the new terrestrial gravity models we have preliminarily used the
recently released EIGEN-GRACE02S gravity model \cite{reigetal04}.
It is important to note that such model represents a long-term
averaged GRACE-only solution (110 days); moreover, the released
sigmas of the spherical harmonic coefficients of the geopotential
are not the mere formal statistical errors, but are calibrated,
although preliminarily. Then, guesses of the impact of the
systematic error due to the geopotential on the measurement of the
Lense-Thirring effect based on this solution should be rather
realistic. However, caution is advised in considering the so
obtained evaluations because of the uncertainties of the
calibration process which affect especially the even zonal
coefficients \ct{rieetal98}.

With regard to the three-elements combination \rfr{ciufform}, it
turns out that the systematic error due to the even zonal
harmonics of the geopotential, according to the variance matrix of
EIGEN-GRACE02S up to degree $l=70$, amounts to 0.2 mas yr$^{-1}$
(Root Sum Square calculation), yielding a 1-sigma $0.4$$\%$ error
in the Lense-Thirring effect. The sum of the absolute values of
the individual errors yields an error of 0.4 mas yr$^{-1}$, i.e. a
1-sigma upper bound of 0.7$\%$ in the Lense-Thirring effect. Of
course, even if the LAGEOS and LAGEOS II data had been reprocessed
with the EIGEN-GRACE02S model, the problems posed by the correct
evaluation of the impact of the non--gravitational perturbations
on the perigee of LAGEOS II would still persist.
%E.g., the authors of \cite{riesetal03} conjecture that, even with
%the best GRACE Earth gravity model, the error in a determination
%of the Lense--Thirring effect using the perigee of LAGEOS II could
%be of the order of 50$\%$--100$\%$, if not larger.

%The
%total systematic error would still remain of the order of
%28$\%$--30$\%$, unless significant improvements in the modeling of
%the non--gravitational perturbations on the perigee of LAGEOS II
%will occur.

A different approach could be followed by taking the drastic
decision of canceling out only the first even zonal harmonic of
the geopotential by discarding at all the perigee of LAGEOS II.
The hope is that the resulting gravitational error is reasonably
small so to get a net gain in the error budget thanks to the fact
that the nodes of LAGEOS and LAGEOS II exhibit a very good
behavior with respect to the non--gravitational perturbations.
Indeed, they are far less sensitive to them than the perigee of
LAGEOS II. Moreover, they can be easily and accurately measured,
so that also the formal, statistical error should be reduced. A
possible observable is  \cite{iorproc04} \eqi\delta\dot\Omega^{\rm
L}+0.546\delta\dot\Omega^{\rm L\ II}\sim 48.2\mu_{\rm
LT}.\lb{iorform}\eqf  A similar proposal can be found in
\cite{riesetal03}, although numerical details are not released
there. According to the variance matrix of EIGEN-GRACE02S up to
degree $l=70$, the residual signal due to the even zonal harmonics
from $l=4$ to $l=70$ is 1.5 mas yr$^{-1}$ (Root Sum Square
calculation), i.e. a 1-sigma 3$\%$ systematic bias in the
Lense-Thirring effect. The sum of the absolute values of the
individual errors yields an upper bound of 1.9 mas yr $^{-1}$,
i.e. a 1-sigma 4$\%$ systematic error. EGM96 would not allow to
adopt \rfr{iorform} because its full covariance matrix up to
degree $l=70$ yields an error of 47.8$\%$ while the error
according to its diagonal part only amounts even to 104$\%$
(1-sigma Root Sum Square calculation), with an upper bound of
177$\%$ (1-sigma sum of the absolute values of the individual
errors). Note also that the combination \rfr{iorform} preserves
one of the most important features of the combination of
\rfr{ciufform} of orbital residuals: indeed, it allows to cancel
out the very insidious 18.6-year tidal perturbation which is a
$l=2,\ m=0$ constituent with a period of 18.6 years due to the
Moon's node and nominal amplitudes of the order of 10$^3$ mas on
the nodes of LAGEOS and LAGEOS II \cite{iorcelmec01}. On the other
hand, the impact of the non--gravitational perturbations on the
combination \rfr{iorform} over a time span of, say, 7 years could
be quantified in 0.1 mas yr$^{-1}$, yielding a 0.3$\%$ percent
error. The results of Table 2 and Table 3 in \cite{ioretal02} have
been used. It is also important to notice that, thanks to the fact
that the periods of many gravitational and non--gravitational
time--dependent perturbations acting on the nodes of the LAGEOS
satellites are rather short, a reanalysis of the LAGEOS and LAGEOS
II data over just a few years could be performed. As already
pointed out, this is not so for the combination \rfr{ciufform}
because some of the gravitational \cite{iorcelmec01} and
non--gravitational \cite{luc01} perturbations affecting the
perigee of LAGEOS II have periods of many years. Then, with a
little time--consuming reanalysis of the nodes only of the
existing LAGEOS and LAGEOS II satellites with the EIGEN-GRACE02S
data it would at once be possible to obtain a more accurate and
reliable measurement of the Lense--Thirring effect, avoiding the
problem of the uncertainties related to the use of the perigee of
LAGEOS II.

The choice of an observational time span of just a few years would
also be helpful in reducing the impact of the secular variations
of the uncancelled even zonal harmonics $\dot J_4$ and $\dot J_6$.
The problem of the impact of the secular variations of the even
zonal harmonics on the proposed measurements of the Lense-Thirring
effect has never been addressed, up to now, in a satisfactorily
way. In \cite{iormor04} it has been claimed, perhaps too
superficially, that the secular variations of the zonals would not
affect the combination \rfr{iorform} because they can be accounted
for by an effective $\dot J_2^{\rm eff}$. In fact, the effective
$J_2$ means a lumped effect that it has not been possible to
separate with one or two satellites. But the individual effects
are still there; they just getted blurred. One can use the lumped
effect to get some insight into the total error in the secular
rates of the zonals, but it tells nothing about the individual
contributions. So, it is not possible to cancel them out in the
combination\footnote{I am grateful to J. Ries for helpful
discussions on this problem.} \rfr{iorform}. For the topic of
$\dot J_l$, which has recently received great attention by the
geodesists' community in view of unexpected variations
of\footnote{Fortunately, any issues concerning $\dot J_2$ do not
affect the combination \rfr{iorform}.} $J_2$, see \cite{cox}. By
assuming $\delta\dot J_4=0.6\times 10^{-11}$ yr$^{-1}$ and
$\delta\dot J_6=0.5\times 10^{-11}$ yr$^{-1}$ \cite{cox}, it turns
out that the 1-sigma error on the combination \rfr{iorform} would
amount to 2.1$\%$ over one year. However, it must be pointed out
that it is very difficult to have reliable evaluations of the
secular variations of the higher degree even zonal harmonics of
geopotential also because very long time series from the various
existing SLR targets are required.
%------------------------------------------------------------
\subsection{Alternative combinations}
In \cite{iormor04} the possibility of using a multisatellite
linear combination including the the nodes of LAGEOS, LAGEOS II,
Ajisai, Starlette and Stella has been investigated. It is, by
construction, insensitive to the first four even zonal harmonics
of the geopotential. On the other hand, the inclusion of the nodes
of the other existing SLR satellites, which orbit at much lower
altitudes than the LAGEOS satellites, introduces, in principle,
much more noise from the higher degree even zonal harmonics which
such combination would be sensitive to. The hope was that the
improvements in the knowledge of just the higher degree even zonal
harmonics from the new GRACE-based solutions would make such an
alternative combination competitive with the two-nodes combination
of \rfr{iorform}. The recent results from EIGEN-GRACE02S rule
neatly out this possibility. Indeed, it turns out that, if, on the
one hand, the problem of the $\dot J_l$ would be greatly reduced,
on the other hand, the 1-sigma Root Sum Square percent error would
be 13$\%$ with an upper bound of 42$\%$ from the sum of the
absolute values.

More favorable and interesting, at least in principle, is the
situation with another node-only combination proposed in
\cite{iordorn04}. It includes the nodes of LAGEOS, LAGEOS II,
Ajisai and the altimeter satellite Jason-1 whose orbital
parameters are similar to those of Ajisai; apart from the LAGEOS
satellites, Ajisai and Jason-1 have the most interesting orbital
configurations, among those of the existing accurately tracked
satellites, for our purposes. The weighing coefficients of the
nodes are 1 for LAGEOS, 0.347 for LAGEOS II, -0.005 for Ajisai and
0.068 for Jason-1; the gravitomagnetic slope is  49.5 mas
yr$^{-1}$. It turns out that the EIGEN-GRACE02S model yields a
systematic gravitational error of 1$\%$ (1-sigma Root Sum Square
calculation) and an upper bound of 2$\%$. Moreover, $\dot J_4$ and
$\dot J_6$ would not affect this combination. However, the
possibility of effectively getting long time series of the node of
Jason-1 should be demonstrated in reality. Finally, dealing
suitably with the non-gravitational perturbations acting on it in
a genuine dynamic way would be a very demanding task.
%------------------------------------------------------------
\section{Conclusions}
When more robust and complete terrestrial gravity models from
CHAMP and GRACE will be available in the near future, the
two-nodes/LAGEOS-LAGEOS II combination \rfr{iorform} could allow
for a measurement of the Lense--Thirring effect with a total
systematic error, mainly due to geopotential, of the order of a
few percent over a time span of some years without the
uncertainties related to the evaluation of the impact of the
non--gravitational perturbations acting upon the perigee of LAGEOS
II. The choice of a not too long observational time span should
also be helpful in keeping the systematic error due to the secular
variations of the even zonal harmonics below the 10$\%$ level.

On the other hand, the obtainable accuracy with the
node-node-perigee combination \rfr{ciufform}, whose error due to
geopotential will remain smaller than that of \rfr{iorform}, is
strongly related to improvements in the evaluation of the
non--gravitational part of the error budget and to the use of time
spans of many years. However, it neither seems plausible that the
error due to the non-conservative forces will fall to the 1$\%$
level nor that a reliable and undisputable assessment of it will
be easily obtained.

Alternative combinations including the orbital data from the other
existing SLR satellites are not competitive with the combination
of \rfr{iorform}.

A combination including also the nodes of the SLR Ajisai and
altimeter Jason-1 satellites, together with the nodes of the two
LAGEOS satellites, would be slightly better from the point of view
of the reduction of the systematic error due to the geopotential,
in particular with regard to the effects of the secular variations
of the even zonal harmonics. However, this gain could be lost due
to the difficulties of dealing with the non-gravitational
perturbations affecting the nodes of Jason-1.

\section*{Acknowledgements}
L. Iorio is grateful to L. Guerriero for his support to him in
Bari and to the GFZ team for the public release of the
EIGEN-GRACE02S gravity model. Thanks also to J. Ries for
stimulating and helpful discussions.


\begin{thebibliography}{xxxxx}

\bibitem{ciuwhe95}
I. Ciufolini and J.A. Wheeler, {\it Gravitation and Inertia},
 (Princeton University Press,
Princeton, 1995)

\bibitem{rufsig03}
L. Stella {\it et al.}, in {\it Nonlinear Gravitodynamics}, edited
by R. Ruffini and C. Sigismondi (World Scientific, Singapore,
2003), p. 235.

\bibitem{nor03}
K. Nordvedt, in {\it Nonlinear gravitodynamics}, edited by R.
Ruffini and C. Sigismondi (World Scientific, Singapore, 2003), p.
35.

\bibitem{sch60}
L. Schiff, Am. J. Phys. {\bf 28}, 340 (1960).

\bibitem{eveetal01}
C.W.F. Everitt {\it et al.}, in {\it Gyros, Clocks,
Interferometers...:Testing Relativistic Gravity in Space}, edited
by C. L$\ddot{\rm a}$mmerzahl, C.W.F. Everitt and F.W. Hehl
(Springer, Berlin, 2001), p. 52.

\bibitem{let18}
J. Lense and H. Thirring,  {\it Phys. Z.} {\bf 19}, 156 (1918),
english
  translation by B. Mashhoon, F. W. Hehl, and D. S. Theiss {\it Gen.
  Relativ. Gravit.} {\bf 16}, 711 (1984).

\bibitem{ciu86}
I. Ciufolini, Phys. Rev. Lett. {\bf 56}, 278 (1986).

\bibitem{ioretal02}
L. Iorio, D. Lucchesi and I. Ciufolini, Class. Quantum Grav. {\bf
19}, 4311 (2002).

\bibitem{ioretal03}
L. Iorio {\it et al.}, Class. Quantum Grav. {\bf 21}, 2139 (2004).

\bibitem{lammetal01}
C. L$\ddot{\rm a}$mmerzahl, H. Dittus, A. Peters and S. Schiller,
Class. Quantum Grav. {\bf 18}, 2499 (2001).

\bibitem{ciuetal98}
I. Ciufolini {\it et al.}, Science {\bf 279}, 2100 (1998).

\bibitem{ciu02}
I. Ciufolini, gr-qc/0209109.

\bibitem{rieetal98}
J. C. Ries, R. J. Eanes and B. D. Tapley, in {\it Nonlinear
Gravitodynamics}, edited by R. Ruffini and C. Sigismondi (World
Scientific, Singapore, 2003), p. 201.

\bibitem{luc01}
D. Lucchesi, Pl. Space Sci. {\bf 49}, 447 (2001).

\bibitem{luc02}
D. Lucchesi, Pl. Space Sci. {\bf 50}, 1067 (2002).

\bibitem{luc03}
D. Lucchesi, {Geophys. Res. Lett.} {\bf 30}, 1957 (2003).

\bibitem{luc04}
D. Lucchesi, {Celest. Mech. Dyn. Astron.} {\bf 88}, 269 (2004).

\bibitem{lucetal04}
D.Lucchesi, {\it et al.}, {Pl. Space Sci.} {\bf 52}, 699 (2004).

\bibitem{lemetal98}
F.G. Lemoine {\it et al.}, NASA/TP-1998-206861, 1998.

\bibitem{iorcelmec03}
L. Iorio, Celest. Mech. Dyn. Astron. {\bf 86}, 277 (2003).

\bibitem{iorcelmec01}
L. Iorio, Celest. Mech. Dyn. Astron. {\bf 79}, 201  (2001).

\bibitem{lucpriv}
D. Lucchesi (private communication).

\bibitem{pav00}
E. Pavlis, in {\it Recent Developments in General Relativity},
edited by R. Cianci, R. Collina, M. Francaviglia and P. Fr${\rm
\acute{e}}$ (Springer, Milan, 2000), p. 217.

\bibitem{riesetal03}
J.C. Ries {\it et al.}, in {\it Proceedings of the 13th
International Laser Ranging Workshop, Washington DC, October 7-11,
2002},
http://cddisa.gsfc.nasa.gov/lw13/lw$\_${proceedings}.html$\#$science

\bibitem{reigetal04}
Ch. Reigber {\it et al.}, J. of Geodynamics, in press, (2004).


\bibitem{iorproc04}
L. Iorio, in {\it Earth Observation with CHAMP. Results from Three
Years in Orbit}, edited by Ch. Reigber, H. L$\ddot{\rm u}$hr, P.
Schwintzer and J. Wickert.  (Springer, Berlin, 2004), p. 187.

\bibitem{iormor04}
L. Iorio and A. Morea, Gen. Rel. Grav. {\bf 36}, 1321 (2004).

\bibitem{cox}
C. Cox {\it et al.}, in {\it Proceedings of the 13th International
Laser Ranging Workshop, Washington DC, October 7-11, 2002},
http://cddisa.gsfc.nasa.gov/lw13/lw$\_${proceedings}.html$\#$science


\bibitem{iordorn04}
L. Iorio and E. Doornbos, Preprint gr-qc/0404062


\end{thebibliography}
\end{document}